\documentclass{article}

\usepackage{arxiv}

\usepackage[utf8]{inputenc} 
\usepackage[T1]{fontenc}    
\usepackage{hyperref}       
\usepackage{url}            
\usepackage{booktabs}       
\usepackage{amsfonts}       
\usepackage{nicefrac}       
\usepackage{microtype}      
\usepackage{lipsum}		
\usepackage{graphicx}
\usepackage{natbib}
\usepackage{doi}
\usepackage{cite}
\usepackage{textcomp}
\usepackage{xcolor}
\usepackage{latexsym,amssymb,amsmath}
\usepackage{longtable}
\usepackage{tikz,pgf}
\usepackage{subfig}
\usepackage{wrapfig}
\usepackage{rotating}
\usepackage{multirow}
\usepackage{lineno,hyperref}
\usepackage{algorithm}
\usepackage{algorithmic}
\usepackage{amsthm}
\usepackage{array}

\input{def.set}

\newcommand{\PreserveBackslash}[1]{\let\temp=\\#1\let\\=\temp}
\newcolumntype{L}[1]{>{\raggedright\arraybackslash}m{#1}}
\newcolumntype{C}[1]{>{\centering\arraybackslash}m{#1}}
\newcolumntype{R}[1]{>{\raggedleft\arraybackslash}m{#1}}

\newtheorem{theorem}{Theorem}

\newtheorem{example}{Example}

\title{Understanding of a brain spatial map \\ based on threshold-free function dendrogramization}

\author{ Hyekyoung Lee$^{*}$ \\
	Seoul National University Hospital\\
	Seoul National University\\
	Seoul 03080 Republic of Korea \\
	\texttt{hklee.brain@gmail.com} \\
	\And
	Hyejin Kang \\
	Seoul National University Hospital\\
	Seoul National University\\
	Seoul 03080 Republic of Korea \\
	\texttt{hkang211@snu.ac.kr} \\
	\And 
	Youngmin Huh \\ 
	Seoul National University\\
	Seoul 03080 Republic of Korea \\
	\texttt{ymin1123@gmail.com} \\
	\And
	Hongyoon Choi \\
	Seoul National University Hospital\\
	Seoul National University\\
	Seoul 03080 Republic of Korea \\
	\texttt{chy1000@snu.ac.kr} \\
	\And 
	Dong Soo Lee  \thanks{Hyekyoung Lee and Dong Soo Lee are corresponding authors. }\\
	Seoul National University Hospital\\
	Seoul National University\\
	Seoul 03080 Republic of Korea \\
	\texttt{dsl@snu.ac.kr} \\
	\And
	for the Alzheimer’s Disease Neuroimaging Initiative \thanks{Data used in preparation of this article were obtained from the Alzheimer’s Disease Neuroimaging Initiative (ADNI) database (adni.loni.usc.edu). As such, the investigators within the ADNI contributed to the design and implementation of ADNI and/or provided data but did not participate in analysis or writing of this report. A complete listing of ADNI investigators can be found at: \url{http://adni.loni.usc.edu/wp-content/uploads/how_to_apply/ADNI_Acknowledgement_List.pdf}} \\ 
}

\renewcommand{\shorttitle}{\textit{arXiv} Template}

\hypersetup{
pdftitle={A template for the arxiv style},
pdfsubject={q-bio.NC, q-bio.QM},
pdfauthor={David S.~Hippocampus, Elias D.~Striatum},
pdfkeywords={First keyword, Second keyword, More},
}

\begin{document}
\maketitle

\begin{abstract}
	Linear matrix factorizations (LMFs) such as independent component analysis (ICA), principal component analysis (PCA), and their extensions, have been widely used for finding relevant spatial maps in brain imaging data. 
The last step of an LMF before interpretation is usually to extract the activated brain regions from the map by thresholding. 
However, it is difficult to determine an appropriate threshold level. 
Thresholding can remove the underlying properties of spatial maps and their features imposed by the model. 
In this study, we propose a threshold-free activated region extraction method which involves simplifying a brain spatial map to a dendrogram through Morse filtration. 
Since a dendrogram is related to the change of clustering structure in Rips filtration, we first show the relationship between the Rips filtration of a graph and the Morse filtration of a function. 
Then, we dendrogramize a spatial map in order to visualize the activated brain regions and the range of their importance in a spatial map. 
The proposed method can be applied to any spatial maps that a user wants to threshold and interpret. 
In experiments, we applied the proposed method to independent component maps (ICMs) obtained from resting-state fMRI data, and the dominant subnetworks obtained by the PCA of a correlation-based functional connectivity of FDG PET Alzheimer's disease neuroimaging initiative (ADNI) data. 
We found that dendrogramization can help to understand a brain spatial map without thresholding. 
\end{abstract}


\label{sec:introduction}
Linear matrix factorization (LMF) models such as principal component analysis (PCA) and independent component analysis (ICA) have been widely used in brain imaging data analysis  \citep{beckmann.2009.ni,bijsterbosch.2017.book,correa.2010.ieeespm}.
An LMF model simply factorizes a brain imaging data matrix into spatial map and feature matrices, in which features usually indicate the relationship between a spatial map and a given task, and the spatial map shows which brain regions are related to the given task.  
The LMF model works in a purely data-driven way. 
However, it can impose a specific property such as correlation, independence, or sparsity onto spatial maps.
A spatial map obtained using LMF can also be considered as a subnetwork in which activated brain regions in the map are strongly connected.  
ICA is a well-known method with which to extract subnetworks as an independent source forming resting state functional connectivity  \citep{calhoun.2012.ieeerbe}. 
Although LMF models do not have a high-level of performance, like deep learning, due to their simplicity and scalability, they have been developed for use in various research fields of neuroscience for discovering new data, such as a new imaging biomarker \citep{beckmann.2009.ni,marchitelli.2018.ni,decheveigne.2019.ni,mohammadi_nejad.2017.tmi,toussaint.2012.ni,woo.2017.nn}.

\begin{figure*}[t]
	\centering
	\includegraphics[width=1\linewidth]{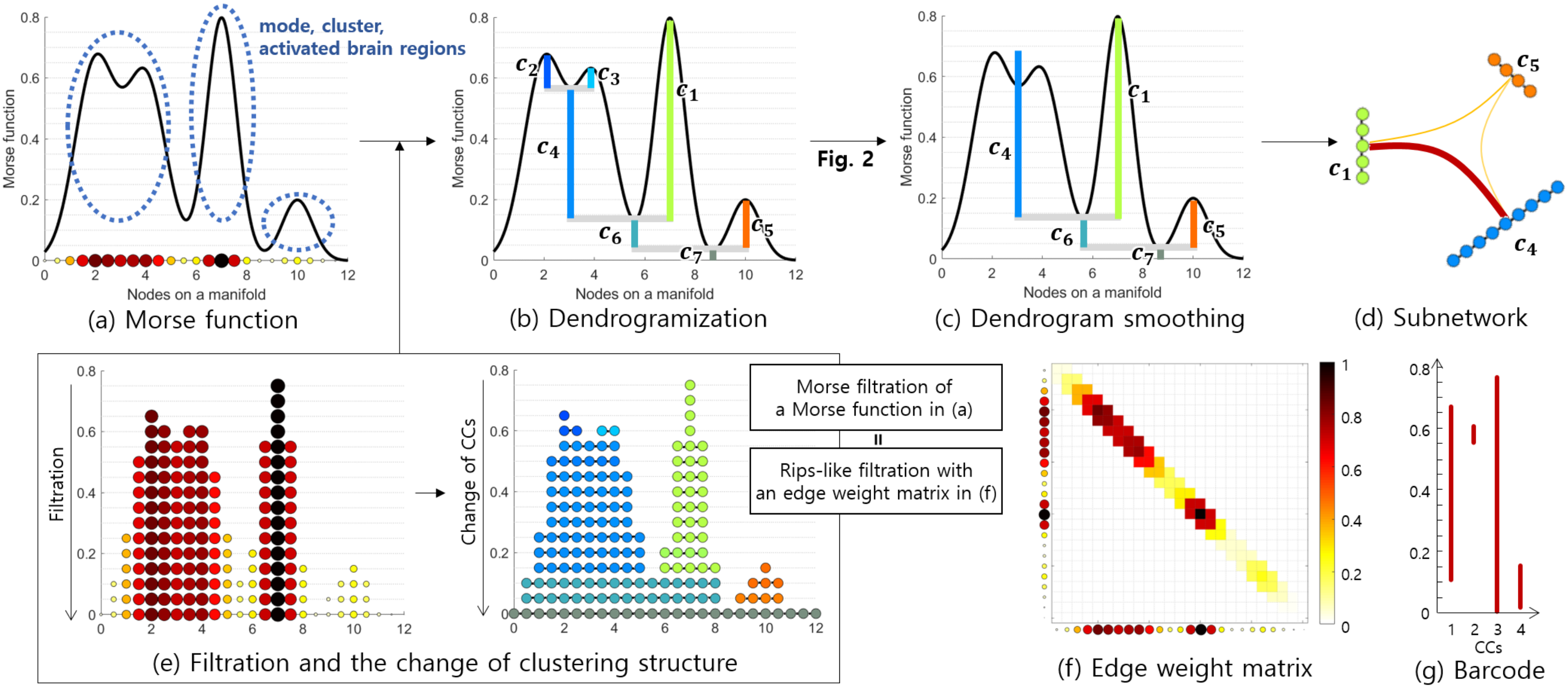}
	\caption{(a) A spatial map (Morse function) on a one-dimensional manifold with 25 points (nodes), $v_{1}, v_{2}, \dots,v_{25}$ from left to right. The goal of this study was to extract clusters or activated regions in the map. (b) Dendrogramization of a spatial map. Dendrogramization is performed by observing the change of clustering structure during filtration in (e). The leaf bars, $c_{2}, c_{3}, c_{1}$ and $c_{5}$ become the activated regions of the Morse function. (c) Dendrogram smoothing. The more detailed procedure is in Fig. \ref{fig:smoothing}. $c_{2}$ and $c_{3}$ are replaced by $c_{4}.$ (d) Subnetwork construction. The spatial map is represented by a network in which nodes are activated regions and edges are determined by the peak value of the activated regions. (e) Filtration and the change of clustering structure at thresholds, $0.75,0.7,\dots,0.$ (f) Edge weight matrix with self-similarity of the Morse function in (a). The Morse filtration of (a) can be thought as the Rips-like filtration of (f). (g) Barcode of CCs.}
	\label{fig:procedure}
	\vspace{-0.25cm}
\end{figure*}

A common procedure in most of the studies on LMF models is to threshold a spatial map for visualization and interpretation. 
While a general linear model (GLM) can estimate the voxel- or cluster-level statistical values of spatial maps based on the hypothesis of a design matrix, LMF cannot be used to directly estimate the significance value in a map \citep{nichols.2012.ni,smith.2009.ni}. 
In the early years, statistical inference in ICA was done by transforming  an independent component map (ICM) to a Z-map, under the assumption that an ICM is normally distributed \citep{mckeown.1998.hbm2}. 
However, since this assumption was not consistent with ICA, 
Beckmann, et. al. proposed a new method based on the Gaussianity of residual errors, instead of that of an ICM \citep{beckmann.2004.tmi}. 
They implemented the proposed method by approximating the histogram of the Z-map of an ICM with a mixture model in which one of mixtures is considered as background noise, and the others are considered as the probabilities of activated brain regions  \citep{beckmann.2004.tmi}. 
Recently, Poppe, et. al. proposed a method to vary thresholds from 0 to 1 after normalizing the maximum value of a spatial map to 1, and choosing a threshold that produced desired results \citep{poppe.2013.cabn,ma.2021.bc}.  
However, we need to ask a fundamental question: ``is it possible to threshold a spatial map extracted using an LMF model?" 
GLM inference measures how well a spatial map fits into a predetermined experimental design, and it is possible to select the parts of a spatial map, called activated regions. 
However, LMF is a data-driven method that estimates spatial maps and corresponding features simultaneously. 
If the parts of a spatial map obtained by LMF are selected by thresholding, the corresponding feature vector should be changed, and eventually, they become not a solution of a given data matrix. 

In this paper, we propose a new threshold-free method for the detection of activated brain regions in a spatial map based on Morse filtration. 
We express the change of the clustering structure of a spatial map during Morse filtration as a tree structure, called a dendrogram \citep{carlsson.2010.jmlr}. 
We call this procedure dendrogramization (Fig. \ref{fig:procedure}). 
A dendrogram usually represents the hierarchical clustering structure of a graph in a metric space. 
To verify the dendrogramization procedure, we show that the Morse filtration of a spatial map on a manifold is related to Rips filtration, if self-distance is allowed \citep{edelsbrunner.2009.book,carlsson.2010.jmlr}. 
Then, we define the leaves of the dendrogram of a spatial map as the activated brain regions of a spatial map. 
Therefore, by dendrogramizing a spatial map, we can see not only the activated brain regions of a spatial map, but also their duration in the map.  
To further simplify a dendrogram by reducing the number of leaves, we also propose a dendrogram smoothing method motivated by persistent homology \citep{edelsbrunner.2009.book,carlsson.2010.jmlr}.   
The higher the peak value of the mode is in a spatial map, the more significant the activated brain region corresponding to the mode is in a spatial map. 

We transform a spatial map to a subnetwork in which nodes are activated brain regions and edges are determined by the peak values of the regions.  
The proposed method can be applied to any spatial map using an LMF model. 
We applied the proposed method to ICMs obtained from resting-state functional connectivity, and dominant spatial maps in a correlation-based functional connectivity obtained by the PCA of FDG PET Alzheimer's disease neuroimaging initiative (ADNI) data.    
Dendrogramization helped to understand the spatial maps by visualizing activated brain regions and the range of their importance in the map, without thresholding.

\section{Proposed methods}
\label{}

\subsection{Morse filtration} 

Suppose that a spatial map $\bs \in \Real^{p \times 1}$ is given.  
This map is a Morse function on a brain manifold, denoted by $s(\bv),$ where $\bv$ is a voxel in a three-dimensional brain imaging space \citep{adler.2010.arxiv}. 
If a three-dimensional brain imaging space is assumed to be a grid coordinate with $v^{\cdot} \in \left\{ 1, \cdots, d_{\cdot} \right\}$ $(\cdot = 1,2,3),$ 
an upper level set at a nonnegative threshold, $\epsilon_{t}$ is defined by:
$$L_{t} (\bs) = \left\{ \bv |  \bs(\bv) \ge \epsilon_{t}, \bv = (v^1,v^2,v^3) \in \Natural^{d_{1} \times d_{2} \times d_{3}} \right\}.$$
We define connected components (CCs, or clusters) in $L_{t}$ by the subset of voxels of which the faces, edges, or corners are touched each other in a three-dimensional space. 
The number of adjacent voxels of each voxel is 26.  
The number of CCs in $L_{t} (\bs)$ is denoted by $\beta_{t},$ which is called the zeroth Betti number. 
The list of CCs in $L_{t} (\bs)$ are denoted by $c_{1}^{t},\dots,c_{\beta_{t}}^{t} \subseteq L_{t},$ which satisfies that $L_{t}(\bs) = c_{1}^{t} \cup \dots \cup c_{\beta_{t}}^{t}$ and $c_{i}^{t} \cap c_{j}^{t} = \emptyset$ for any $i,j \in \left\{ 1, \cdots, \beta_{t} \right\}.$ 
Given the sequence of thresholds, $\epsilon_{1} \ge \epsilon_{2} \ge \cdots \ge \epsilon_{T} \ge 0,$ the sequence of level sets satisfies the nested property such that $L_{1}(\bs) \subseteq L_{2}(\bs) \subseteq \dots \subseteq L_{T}(\bs).$ 
The nested sequence of upper level sets is called Morse filtration  \citep{adler.2010.arxiv,edelsbrunner.2009.book}. 
A CC that lasts from $\epsilon_{t}$ to $\epsilon_{t+\tau-1}$ such that $c^{t} \subseteq c^{t+1} \subseteq \cdots \subseteq c^{t+\tau-1}$ is called persistent CC (PCC). 
The thresholds $\epsilon_{t}$ and $\epsilon_{t+\tau-1}$ are called the birth and death thresholds of the PCC, respectively. 
Their difference, $\tau = \epsilon_{t+\tau-1} - \epsilon_{t}$ is defined as the duration of the PCC. 

\begin{figure*}[t]
	\centering
	\includegraphics[width=1\linewidth]{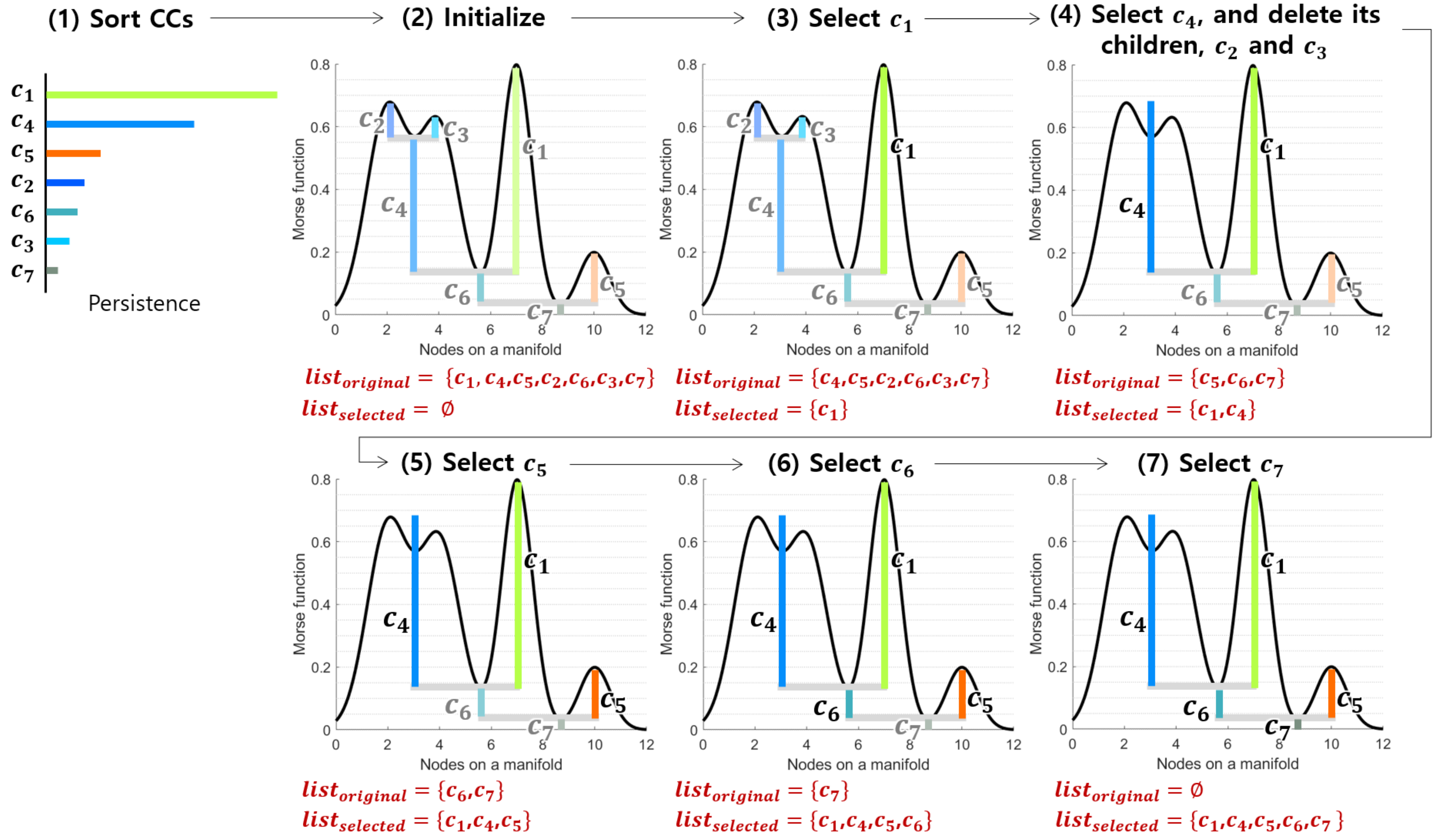}
	\caption{Dendrogram smoothing of Fig. \ref{fig:procedure} (b) in Example \ref{ex:2}.}
	\label{fig:smoothing}
\end{figure*}

\subsection{Hierarchical clustering structure in a Morse function}
\label{sec:dendrogram}

During Morse filtration, new voxels are added at every threshold. 
The new voxels originally become a new PCC. 
Some of them merge with other adjacent PCCs and disappeared without a trace, becoming a new PCC. 
For convenience, we omit the process of becoming PCCs in this case, and simply divide the change of current clustering structure by the new voxels into three cases as follows:  
\begin{enumerate}
	\item[(1)] creating a new CC, if some of them are isolated, 
	\item[(2)] being connected into the existing CC, if some of them are adjacent to the existing CC, and 
	\item[(3)] merging more than two existing CCs into a new CC, if some of the new voxels are adjacent with the existing CCs.  
\end{enumerate} 

The clustering structure is changed by cases (1) and (3). 
In case (2), only the size of the existing PCC, which is the number of voxels in the PCC, is expanded, but there is no change in the clustering structure.    
Case (1) creates a new PCC. 
In case (3), the change of PCC structure is slightly different between traditional persistent homology and hierarchical clustering dendrograms. 
For example, suppose that two PCCs, $c_{i}$ and $c_{j},$ are merged into $c_{k}$ at $\epsilon_{t}.$ 
In traditional persistent homology, if $c_{i}$ appears earlier than $c_{j}$ during the filtration, $c_{i}$  continues ($i=k$), and $c_{j}$ disappears. 
However, in a hierarchical clustering dendrogram, both $c_{i}$ and $c_{j}$ disappear, and a new PCC $c_{k}$ starts at $\epsilon_{t}.$ 
In this study, we will represent the change of connected structure during Morse filtration by hierarchical clustering dendrograms. 
The new PCC is called a parent, and the combined PCCs are called children (children $\subset$ parent). 

\subsection{Relationship between Morse filtration and Rips filtration}
\label{sec:Rips_Morse}

Rips filtration is defined on a metric space with point cloud data and a distance between points \citep{carlsson.2010.jmlr}. 
If a weighted graph is in a metric space, 
the topological change of a given weighted graph during Rips filtration is related to the single linkage hierarchical clustering structure of the graph, and it can be expressed as a dendrogram \citep{carlsson.2010.jmlr,lee.2012.ieeemi}. 
Therefore, if the relationship between Rips filtration and Morse filtration can be found, it will be clearer whether a Morse function can be expressed as a dendrogram or not. 

If a space in which a Morse function is represented as a grid coordinate, the space looks like a lattice graph in which voxels are nodes and adjacent voxels are connected by edges.  
Suppose that an edge weight (similarity) matrix of a lattice graph is given by $\bW = [w_{ij}] \in \Real^{p \times p},$ where 
\be 
\label{eq:edgeweight}
w_{ij} = \left\{ \begin{array}{ll} s_{i}, & \mbox{if $i = j,$} \\ 
	\min (s_{i},s_{j}), & \mbox{if $\bv_{i}$ and $\bv_{j}$ are adjacent, } \\ 
	0, & \mbox{otherwise,} \end{array} \right. 
\ee 
where $s_{i} = s(\bv_{i})$ for $i = 1,\dots p.$ 

\begin{theorem} 
Suppose that a node set, $V = \left\{ v_{1}, \dots,v_{p}\right\}$ and a nonnegative function $s:V \rightarrow \Real^{+}$ are given. A distance matrix between nodes is defined by $\bD = [d_{ij}] \in \Real^{p \times p}$ with   
\be 
\label{eq:distance}
d_{ij} = \left\{ \begin{array}{ll} d_{i} = 1/s_{i}, & \mbox{if $i = j,$} \\ 
	\max(d_{i},d_{j}), & \mbox{if $\bv_{i}$ and $\bv_{j}$ are adjacent, } \\ 
	d_{\max}, & \mbox{otherwise,} \end{array} \right. 
\ee 
where $d_{\max}$ is a large enough constant. 
The nodes $v_{1}, \dots,v_{p}$ with a distance matrix $\bD$ are in a partial metric space. (Proof in Appendix \citep{bukatin.2009.amm})
\end{theorem} 

In a metric space, every point (node) has a distance of 0 from itself, so the self-distance of a node is always 0. 
Therefore, all nodes exist from the beginning of Rips filtration at threshold 0, and the number of CCs at 0 is the same as the number of nodes. 
However, in a partial metric space, a self-distance is not zero as shown in (\ref{eq:distance}). 
If we perform Rips-like filtration on a partial metric space with $V$ and $\bD,$ different nodes appear at different thresholds. 

\begin{theorem} 
Given a node set $V = \left\{ v_{1}, \dots,v_{p}\right\}$ with a nonnegative function $\bs = [s_{i}=s(v_{i})] \in \Real^{p \times 1}$ and a distance matrix $\bD = [d_{ij}] \in \Real^{p \times p}$ in (\ref{eq:distance}), the Morse filtration of $\bs$ with threshold $\epsilon_{1} \ge \cdots \ge \epsilon_{T} \ge 0$ is the same as the Rips-like filtration of $\bD$ with threshold $1/\epsilon_{1} \le \cdots \le 1/\epsilon_{T} \le d_{max}.$
\end{theorem} 
\emph{Sketch of proof:} 
Nodes are added in ascending order of $s_{i}$ in Morse filtration, and in descending order of $d_{i}$ in Rips-like filtration. 
Since $d_{i} = 1/s_{i},$ the order of added nodes is the same. 
If an adjacency matrix is $\bA = [a_{ij}] = (\bD > 0)$ in (\ref{eq:distance}), $a_{ij}=1$ when two nodes are adjacent on a brain manifold, and $0$ when they are not. 
If $v_{i}$ and $v_{j}$ are adjacent, an edge between $v_{i}$ and $v_{j}$ is added at $\max(d_{i}=1/s_{i},d_{j}=1/s_{j})$ in Rips-like filtration. 
This is the same as the moment at which $v_{i}$ and $v_{j}$ appear at $\min(s_{i},s_{j})$ in Morse filtration.

The hierarchical clustering structure of Morse filtration is the same as that of Rips-like filtration.

\subsection{Dendrogramization and active mode detection of a spatial map} 

A dendrogram is a visualization tool of the hierarchy of PCCs in a tree form on a two-dimensional plane, as shown in Fig. \ref{fig:procedure} (b). 
The vertical and horizontal axes in a dendrogram represent the threshold and the index of the PCCs, respectively.  
In a dendrogram, a PCC is represented by a vertical bar, which starts at the birth threshold and ends at the death threshold.  
Children and parents are connected by a horizontal bar.  
The last bars of the tree, without any children are called leaves.
We consider the leaves of dendrogram as the modes of a Morse function, and the modes of a Morse function as the activated brain regions in a spatial map. 

\begin{example}
	\label{ex:1}
	Fig. \ref{fig:procedure} (a) shows an example of a 1-dimensional Morse function, $\bs,$ with 25 nodes, denoted by $v_1, \cdots, v_{25}$ from 0 to 12 on the horizontal axis.  
	In this example, a voxel has two neighbors on the left and right except for voxels at both ends. 
	Morse filtration is performed for thresholds, $0.75 > 0.7 > \cdots > 0,$ and PCCs, $c_{1}, \dots, c_{7}$ are found as shown in Fig. \ref{fig:procedure} (e).  
	A dendrogram in (b) shows the hierarchical clustering structure of a Morse function $\bs$ in (a) as well as that of nodes with an edge weight matrix in (f). 
	Note that the traditional Morse filtration finds 4 PCCs as shown in the barcode in (g).  
	The leaves of dendrogram in (b), $c_{2}$, $c_{3}$, $c_{1}$, and $c_{5},$ become the activated regions of the map such that $c_{2} = \left\{v_{5}, v_{6}\right\},$ $c_{3} = \left\{v_{8},v_{9}\right\},$ $c_{1} = \left\{v_{13}, \cdots, v_{17}\right\},$ and $c_{5} = \left\{v_{19}, \cdots, v_{22}\right\}.$ 
\end{example}

\subsection{Dendrogram smoothing}
\label{sec:dendrogram}

We propose a dendrogram smoothing method to simplify a spatial map by reducing the number of modes in the map, by restricting the size and duration of a PCC. 
First, dendrogram smoothing was performed by deleting PCCs of which size is smaller than a predefined minimum size. 
During deletion, if all children are deleted, their parent PCC absorbs all of those children, and the birth threshold of the parent PCC is replaced by the maximum birth threshold of all deleted children.  
After the deletion, if there is a PCC which has an only child, the child is deleted, even if they are larger than the minimum size. 
The children of the child become the children of the PCC which had the deleted child.
The birth threshold of the PCC, which had the deleted child, is also replaced by the birth threshold of the deleted child. 
This procedure is repeated until the size of all PCCs is larger than the minimum size.    
The minimum size of a PCC is determined by the user, and determines the resolution of a smoothed dendrogram.

Second, dendrogram smoothing was performed by choosing PCCs having longer duration, under the assumption of persistent homology that the longer the duration of a PCC, the better the underlying clustering structure of a Morse function is reflected \citep{adler.2010.arxiv,edelsbrunner.2009.book}. 
We initialize an original list of all PCCs sorted in descending order of their duration, and a selected list by an empty set.  
The PCC with the longest duration is selected from the original list.  
Then, we put it on the selected list by updating its birth threshold by the maximum peak value of voxels in the PCC, and remove the selected PCC and all of its descendants in the original list. 
Again, the PCC with the longest duration is selected from the original list. 
If all its children are removed from the original list and are in the selected list, the PCC is also removed from the original list and put into the selected list. 
Otherwise, the PCC is moved to the selected list by updating its birth threshold by the maximum peak value of its voxels, and removed from the original list with all of its descendants, as above. 
This procedure is repeated until the original list is empty.  
Finally, we reconstruct a dendrogram consisting of PCCs in the selected list, and call it a smoothed dendrogram. 
We call this procedure dendrogram smoothing. 
The leaves of a smoothed dendrogram can be thought as the modes of a smoothed Morse function, and we consider them to be the activated brain regions of a simplified brain spatial map.

\begin{example}
\label{ex:2}
We apply dendrogram smoothing method to Fig. \ref{fig:procedure} (b). 
In this example, the minimum size of a PCC is 0. 
In Fig. \ref{fig:smoothing} (1), we first sort bars in descending order of duration.  
Then, we initialize two lists of PCCs, $list_{original}$ and $list_{selected},$ as shown in (2). 
In (3), the longest green bar $c_{1}$ is selected, and it moves to $list_{selected}.$ 
In (4), the longest blue bar $c_{4}$ is selected, but it has two children, $c_{2}$ and $c_{3}.$ 
The $c_{2}$ and $c_{3}$ are absorbed to $c_{4}$ by updating the birth threshold of $c_{4}.$ 
The $c_{2},$ $c_{3},$ and $c_{4}$ are removed from $list_{original},$ and only $c_{4}$ moves to $list_{selected}.$ 
In (5), the longest orange bar $c_{5}$ is selected, and it moves from $list_{original}$ to $list_{selected}.$  
In (6) and (7), the next longest blue-green and khaki bars, $c_{6}$ and $c_{7}$ are selected, respectively. 
Since their children are not in $list_{original}$ at the step (6) and (7), it just moves from $list_{original}$ to $list_{selected}.$ 
Finally, $list_{original}$ becomes empty, and the smoothing procedure is finished. 
The smoothed dendrogram is shown in Fig. \ref{fig:procedure} (c).
The blue, green, and orange bars become the activated brain regions of a given spatial map (Morse function) such that $c_{4} = \left\{v_{3}, \cdots, v_{11}\right\},$ $c_{1} = \left\{v_{13}, \cdots, v_{17}\right\},$ and $c_{5} = \left\{v_{19}, \cdots, v_{22}\right\}.$ 
\end{example}

\subsection{Relationship between a dendrogram and a smoothed dendrogram}

Suppose that the hierarchy of PCCs having $v_{i}$ and $v_{j}$ in a dendrogram is as follows:  
\begin{center} 
	\begin{tikzpicture}
		\node (a) at (0,1) {$v_{i} \in c_{1}^{i}$};  
		\node (b) at (1.5,1) {$c_{2}^{i}$} edge [<-] (a);  
		\node (c) at (2.7,1) {$\cdots$} edge [<-] (b); 
		\node (d) at (3.9,1) {$c_{l}^{i}$} edge [<-] (c); 
		\node (e) at (0,0) {$v_{j} \in  c_{1}^{j}$}; 
		\node (f) at (1.5,0) {$c_{2}^{j}$} edge [<-] (e); 
		\node (g) at (2.7,0) {$\cdots$} edge [<-] (f); 
		\node (h) at (3.9,0) {$c_{m}^{j}$} edge [<-] (g); 
		\node (i) at (5.7,0.5) {$c_{1}^{ij} \ni v_{i}, v_{j}$} edge [<-] (d) 
		edge[<-] (h); 
	\end{tikzpicture}
\end{center} 
where $c_{l}^{i}$ and $c_{m}^{j}$ are PCCs having $v_{i}$ and $v_{j},$ respectively, and $c_{1}^{ij}$ are a CC where $v_{i}$ and $v_{j}$ meet for the first time. 
We assume that $l \ge m.$ 
If $c_{1}^{ij}$ has the longest duration, $c_{1}^{i}, \dots, c_{l}^{i}$ and $c_{1}^{j}, \dots, c_{m}^{j}$ are removed by dendrogram smoothing. 
If we denote the birth threshold of $c$ in a dendrogram and a smoothed dendrogram by $b_{original} (c)$ and $b_{smoothed} (c),$ respectively, the birth threshold of $c_{1}^{ij}$ is updated after dendrogram smoothing by:  
\bee
b_{smoothed} (c_{1}^{ij}) &=& \max(b_{original}(c_{1}^{i}),b_{original}(c_{1}^{j})). 
\eee
Therefore, the difference between original and smoothed dendrograms is:  
\bee 
&&|b_{smoothed}(c_{1}^{ij}) - b_{original}(c_{1}^{ij})|  \\ 
 &=&|\max(b_{original}(c_{1}^{i}),b_{original}(c_{1}^{j})) - b_{original}(c_{1}^{ij})| \\
 &=& \max(\tau(c_{1}^{i})+\cdots+\tau(c_{l}^{i}),\tau(c_{1}^{j})+\cdots+\tau(c_{m}^{j})) \\
 &\le& l \tau(c_{1}^{ij}), 
 \eee
 where $\tau(c)$ is the duration of $c.$

\subsection{Spatial map estimation by LMF}

Given a data matrix $\bX \in \Real^{n \times p}$ with $n$ participants or time points and $p$ voxels, an LMF model decomposes $\bX$ into a spatial map matrix $\bA \in \Real^{n \times k}$ and a feature matrix $\bS \in \Real^{k \times p}$ such that $\bX \approx \bA \bS$, where $k$ is the predefined number of spatial maps. 
The column vector of $\bA,$ denoted by $\ba \in \Real^{n \times 1},$ and the row vector of $\bS,$ denoted by $\bs^{\top} \in \Real^{1 \times p},$ are a feature and spatial map, respectively.  
ICA finds spatial maps that are an independent source generating $\bX,$ and PCA finds a spatial map that consists of maximally covariated brain regions in $\bX.$

If the column vector of $\bX$ is centered with zero mean, PCA can be written in an optimization problem such as 
\be 
\label{eq:pca}
\begin{array} {lcl}
	\argmin_{a,s} \parallel \bX - \ba \bs \parallel^{2}_{2} & \Leftrightarrow & \argmax_{s} \bs \bX^{\top} \bX \bs^{\top}  \\
	\mbox{subject to } \bs \bs^{\top} = 1 & &\mbox{subject to } \bs \bs^{\top} = 1.
\end{array}
\ee  
In the right part of (\ref{eq:pca}), $\bX^{\top} \bX \in \Real^{p \times p}$ is a covariance matrix of $\bX$ in standard PCA.  
However, if the column vector of $\bX$ is centered and normalized with zero mean and length one, $\bX^{\top} \bX$ becomes a correlation matrix, which is widely used as a representation of functional brain connectivity. 
If $\bX$ is the data matrix of a group, $\bs$ represents the dominant subnetwork of the functional connectivity of the group. 
If $\bX$ is the data matrix of all groups, for example, normal controls (NCs), stable mild cognitive impairments (sMCIs), progressive MCIs (pMCIs), and Alzheimer's disease (AD) patients, $\bs$ represents the dominant subnetwork which consists of maximally correlated brain regions over groups, that is, changes during the Alzheimer's disease progression. 
If the number of voxels, $p,$ is too large to directly compute  $\bX^{\top} \bX,$ we can reduce the computational complexity by the singular value decomposition of $\bX$ instead of the PCA of a correlation matrix. 
In this way, although we cannot observe a voxel-level correlation matrix directly, we can find a subnetwork mainly dominated in a correlation-based functional connectivity by basic LMF method, PCA.

\section{Results}

\subsection{Toy example: changing the minimum cluster size} 

\begin{figure*}[t]
	\centering
	\includegraphics[width=1\linewidth]{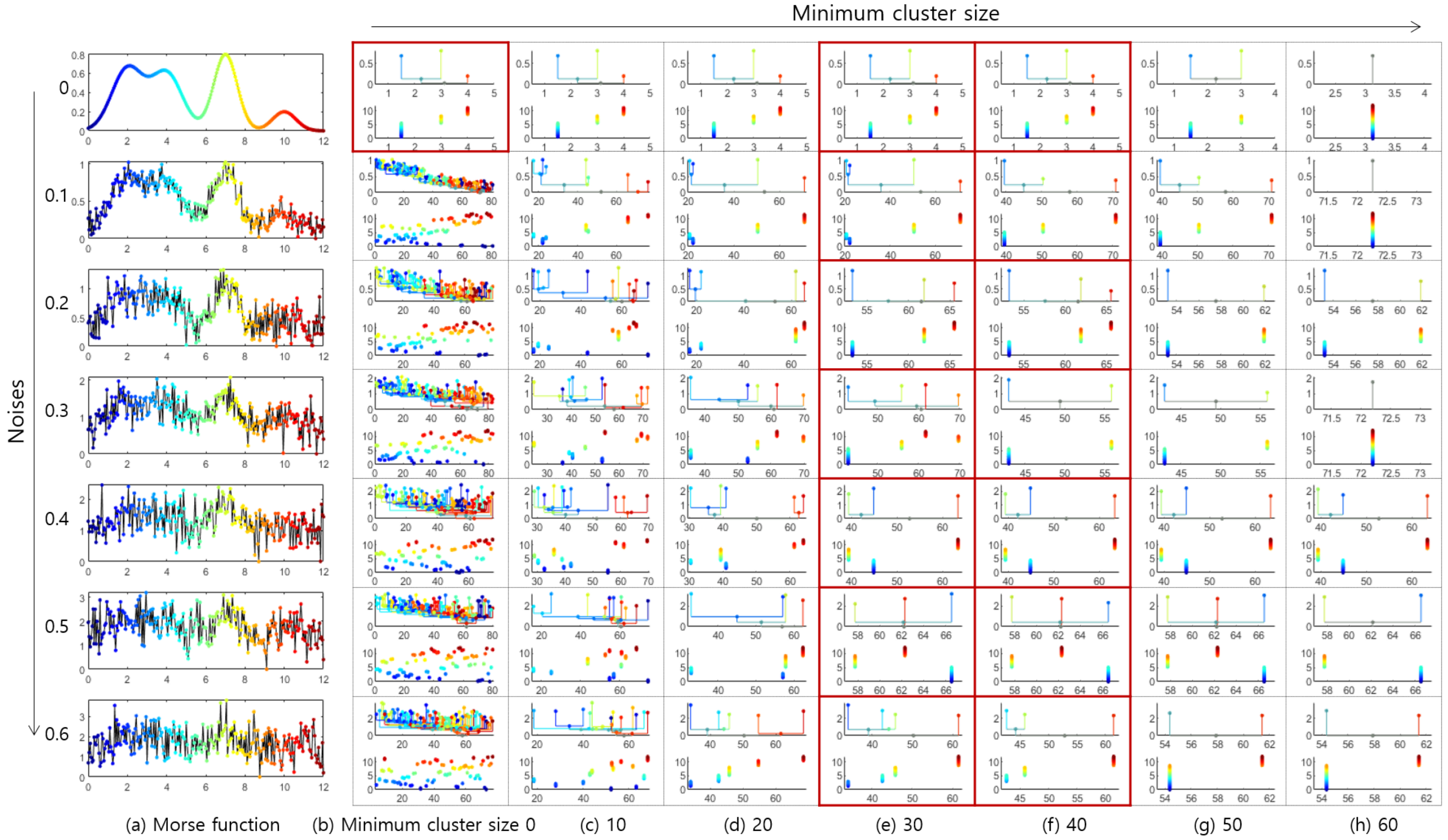}
	\caption{Dendrograms and activated regions of Example \ref{ex:1} with noises. (a) Morse functions of Example \ref{ex:1} by varying the magnitude of noises, $0, 0.1, \dots, 0.6$ from top to bottom. Dendrograms and activated brain regions of the Morse function on the left with the minimum cluster size (b) 0, (c) 10, (d) 20, (e) 30, (f) 40, (g) 50, and (h) 60. In each figure, the upper panel shows a dendrogram and the lower panel shows activated regions corresponding to the leaves of the above dendrogram. In the lower panel, the horizontal and vertical axes correspond to the horizontal axis of the above dendrogram and the horizontal axis of the Morse function in (a), respectively. The color of a bar in a dendrogram was determined by the average color of points in an activated region corresponding to the bar. }
	\label{fig:toyexample_noise}
	\vspace{-0.25cm}
\end{figure*}

We first observed the change of smoothed dendrogram by changing the signal-to-noise ratio and the minimum cluster size. 
We generated a Morse function $\bs = [s_{i}]$ for $i = 1,\dots,241$ by mixing four Gaussian distributions as follows: 
\bee
s_{i} &=& 1.3 \calN(v_{i}; 2,0.8)+1.2 \calN(v_{i}; 4,0.8) \\ 
&& +1.2 \calN(v_{i}; 7,0.6) +0.3 \calN(v_{i}; 10,0.6) + w \cdot e,
\eee
where $\calN(v_{i}; \mu,\sigma)$ is a Gaussian distribution of a variable $v_{i} \in \left\{ 0,0.05,\cdots,12 \right\},$ with mean $\mu$ and standard deviation $\sigma,$ $e$ is Gaussian noise following $\calN(e; 0,1),$ and  $w$ is a parameter controlling the magnitude of noise. 
The generated Morse function, $\bs,$ was shown in Fig. \ref{fig:toyexample_noise} (a) for $w = 0,0.1,0.2,\cdots,0.6$ from top to bottom. 
The 241 nodes each had its own color from blue through green to red. 
We obtained the dendrogram and activated regions of each Morse function by varying the minimum size of a cluster, $0, 10, 20, \dots, 60$ in (b-h).  

The first row in (b-h) shows the smoothed dendrograms and the activated regions of the true Morse function without noise, but with different minimum sizes. 
As the minimum size increased, the small clusters, like the red nodes, disappeared from the dendrogram, as shown in (g,h). 
Although the noise levels of the functions were different, the activated regions obtained by the smoothed dendrograms were somewhat similar to each other at the minimum cluster size 30 in (e).  
However, because of the noise, the birth and death thresholds of the activated regions (clusters) were different from those of the true dendrogram without noise.  

\begin{figure*}[t]
	\centering
	\includegraphics[width=1\linewidth]{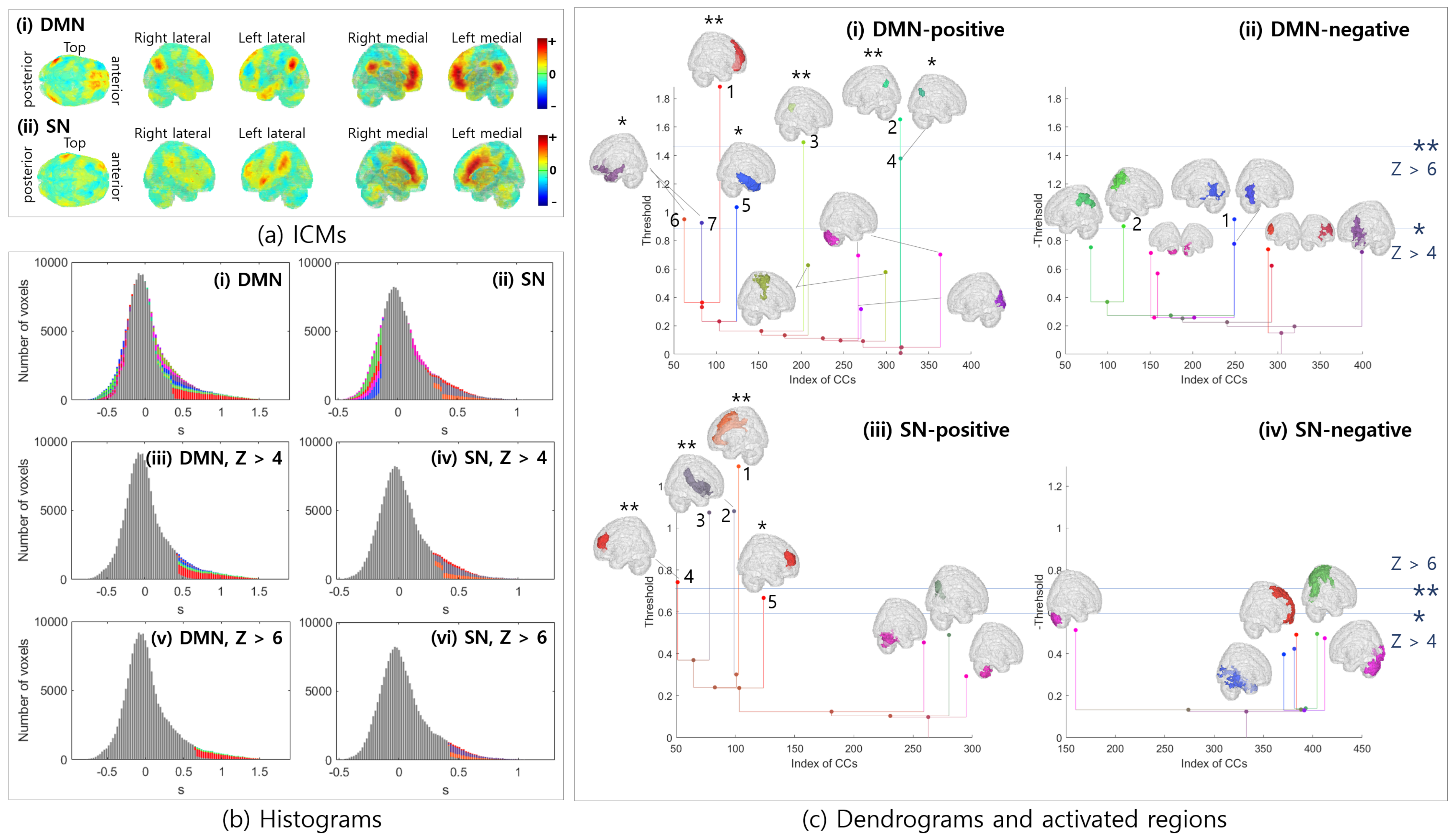}
	\caption{(a) ICMs, (i) DMN and (ii) SN (bottom). (b) Histogram of DMN and SN. The colored region in a histogram corresponds to voxels in activated brain regions obtained by (i,ii) the proposed method, (iii,iv) $Z>4,$ and (v,vi) $Z>6.$ The color is the same as the color of the corresponding activated brain regions. The gray region represents voxels that were not selected as any activated brain regions. (c) Dendrograms of (i,ii) DMN and (iii,iv) SN in (a). The positive and negative parts are on (i,iii) and (ii,iv), respectively. Activated brain regions are plotted on the corresponding leaf bar. Activated brain regions that survived $Z > 4$ and $Z>6$ are represented by $\ast$ and $\ast \ast,$ respectively. Detailed brain regions are listed in Table \ref{table:ICMs}. }
	\label{fig:ICMs}
	\vspace{-0.25cm}
\end{figure*}

\subsection{ICMs of resting-state fMRI} 


\subsubsection{Data set and ICA} 
Resting-state fMRI data was measured from 20 healthy subjects (male = 4; female = 16, mean age = 26.45 $\pm$ 5.20, mean education year = 15.55 $\pm$ 2.77) when they closed their eyes (TR = 3000ms, TE = 30ms, flip angle = 90°, FOV = 240 $\times$ 240mm, matrix size 128 $\times$ 128, 45 axial slices with 3mm thickness, voxel size = 1.9 $\times$ 1.9 $\times$ 3.0mm$^{3}$, resulting 180 volumes). 
The imaging data was preprocessed using Statistical Parametric Mapping (SPM, www.fil.ion.ucl.ac.uk/spm/) and FMRIB Software Library (FSL, fsl.fmrib.ox.ac.uk). 
The preprocessing procedure was as follows: (1) discarding five volumes, (2) repairing using ArtRepair Software (https://cibsr.stanford.edu/tools/human-brain-project/artrepair-software.html) \citep{mazaika.2005.hbm}, (3) slice timing and motion artifacts corrections, (4) co-registeration to anatomical T1 weighted images and normalization to Montreal Neurological Institute (MNI) space, (5) smoothing and intensity normalization, (6) wavelet despiking \citep{patel.2016.ni}, (7) regressing out of white matter and CSF signals, and six motion parameters, and (8) bandpass filtering (0.01Hz-0.1Hz).  
ICA was performed by using Multivariate Exploratory Linear Decomposition into Independent Components (MELODIC). 
Among ICMs, we manually selected a default mode network (DMN) and a salience network (SN), which are plotted in Fig. \ref{fig:ICMs} (a). 
We also estimated the threshold values of $Z > 4$ and $Z > 6$ in DMN and SN by using the Post-Stats method in MELODIC. 

\subsubsection{Dendrogramization and activated regions} 
\label{sec:ICAresult}

A spatial map, $\bs,$ usually has positive and negative parts, denoted by $\bs_{+} > 0$ and $\bs_{-} < 0,$ respectively.  
Therefore, we dendrogramized $\bs_{+} > 0$ and $-\bs_{-} > 0$ separately, and showed the smoothed dendrograms of the positive and negative parts separately on the left and right of Fig. \ref{fig:ICMs} (c), respectively. 
We chose a minimum cluster size of 800. 
The activated brain regions of the ICMs were plotted on the corresponding bars in a dendrogram. 
The color of the voxels was set according to their location in the brain, red for frontal, green for parietal, blue for temporal, violet for occipital, green and orange for subcortical, and purple for cerebellar regions. 
The color of the brain regions and corresponding bars was determined by averaging the colors of voxels in the brain region. 
In the dendrograms, the threshold values of $Z > 4$ (*) and $Z > 6$ (**) were plotted as the blue dashed lines.

The detailed names of the activated brain regions in Fig. \ref{fig:ICMs} (c) are given in Table \ref{table:ICMs}. 
In the dendrogram in Fig. \ref{fig:ICMs} (c-i), there are three big gaps between the birth values of the activated brain regions. 
The first one was between the fourth and fifth regions (1.38 and 1.04), the second one was between the first and second regions (1.88 and 1.65), and the third one was between the seventh and eighth regions (0.93 and 0.70). 
The second and fourth activated regions were the left and right ANG, respectively. 
If the spatial map was thresholded at $Z>6,$ only the left ANG was detected, and the right ANG was missed by a narrow margin. 
In ICM, the negative part was usually rarely considered. 
In our DMN, when thresholding the map at the same threshold $Z>4$ in the positive part, two activated regions, MTG and ITG (blue) and right PCNU (green), survived (Fig. \ref{fig:ICMs} (c-ii)). 
    
In the dendrogram in Fig. \ref{fig:ICMs} (c-iii), there are also three big gaps between the third and fourth activated regions (1.07 and 0.74), the first and second regions (1.29 and 1.08), and the fifth and sixth regions (0.67 and 0.49). 
If the threshold level was selected by $Z>6$ (**), the right MFG and SFG would be missed. 
The histograms of DMN and SN are plotted in Fig. \ref{fig:ICMs} (b), and they show the distribution of the activated brain regions selected by the proposed method, $Z>4,$ and $Z>6$ from top to bottom according to the colored regions.

\begin{figure*}[t]
	\centering
	\includegraphics[width=0.95\linewidth]{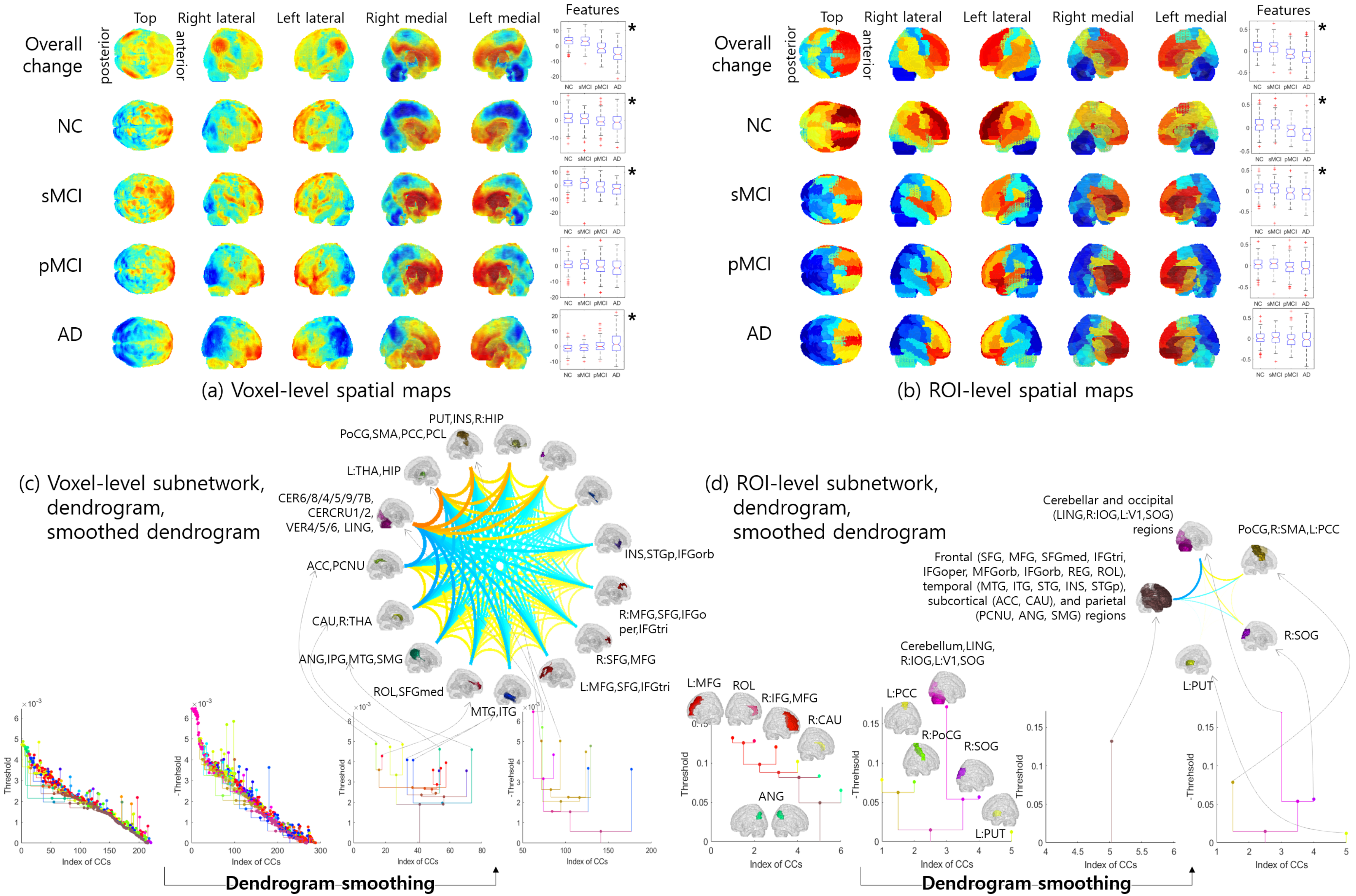}
	\caption{(a,b) Dominant spatial maps representing the overall changes during disease progression, and those representing NC, sMCI, pMCI, and AD in voxel- and ROI-level functional connectivity. Each spatial map is plotted in the top, right and left lateral, and the right and left medial views from left to right. The boxplot of the corresponding features for NC, sMCI, pMCI, and AD are shown on the right. (c,d) Subnetworks (top), dendrograms (bottom, left), and smoothed dendrograms (bottom, right) of the voxel- and ROI-level dominant spatial maps of the overall change in (a,b). The positive and negative parts of a spatial map are dendrogramized on the left and right, respectively. In a circular subnetwork, activated brain regions corresponding to the leaf bars in a dendrogram became nodes. If two nodes are within the same positive or negative parts, they are connected by red, orange, or yellow lines. Otherwise, they are connected by blue or sky blue lines. The line width and the darkness of the line color is proportional to the peak values of the connected nodes.}
	\label{fig:ADNI}
	\vspace{-0.25cm}
\end{figure*}

\subsection{Subnetwork extraction from FDG PET ADNI data} 

\subsubsection{Data set} 

Data used in the preparation of this article were obtained from the Alzheimer’s Disease Neuroimaging Initiative (ADNI) database (adni.loni.usc.edu). The ADNI was launched in 2003 as a public-private partnership, led by Principal Investigator Michael W. Weiner, MD. The primary goal of ADNI has been to test whether serial magnetic resonance imaging (MRI), positron emission tomography (PET), other biological markers, and clinical and neuropsychological assessment can be combined to measure the progression of mild cognitive impairment (MCI) and early Alzheimer’s disease (AD).
The FDG PET images in the ADNI dataset consisted of 182 NCs, 92 sMCIs, 78 pMCIs, and 138 AD patients (age: $73.7\pm5.9$, range $56.1–90.1$). 
The FDG PET images were measured from 30 to 60 min, and averaged over all frames. 
The images were spatially normalized to MNI space using SPM software (SPM8, www.fil.ion.ucl.ac.uk/spm). 
More details are given in \citep{choi.2018.bbr}.
The dimensions of the brain image data were $79 \times 95 \times 68.$ 
We selected $182,363$ voxels which were labeled by a second version of automated anatomical labeling atlas (AAL2) \citep{rolls.2015.ni}. 

\subsubsection{Dominant subnetwork of Alzheimer's disease progression} 
\label{sec:ADNIresult}

We obtained the dominant spatial maps of all groups, NC, sMCI, pMCI, and AD by the PCA of voxel- and ROI-level correlation matrices in (\ref{eq:pca}). 
The voxel- and ROI-level dominant spatial maps are plotted in Fig. \ref{fig:ADNI} (a,b).
We also estimated features by projecting the spatial maps to the data matrix of all groups, and the boxplots of the features were plotted for four groups in (a,b).  
The asterisk (*) on the right indicates that the spatial map of overall change in the first row could significantly discriminate between the four groups (Kruskal-Wallis test, $p<.005$), and that of NC, sMCI, pMCI, and AD in the next four rows could significantly discriminate the group from the other groups (Wilcoxon rank sum test, $p < .005$).

Fig. \ref{fig:ADNI} (c,d) showed the original and smoothed dendrograms, and subnetworks of the voxel- and ROI-level dominant spatial maps of overall change, respectively, in the first row in (a,b). 
We set the minimum cluster size to 400 for dendrogram smoothing.
In the voxel-level subnetwork in (c), the positive part consisted of the ACC, PCNU, ANG, prefrontal region (ROL, SFGmed, MFG, SFG, IFGtri), and MTG and ITG with caudate (CAU), which are related to the DMN. 
The negative part mainly consisted of the cerebellum, hippocampus (HIP), and motor cortex (postcentral gyrus (PoCG), supplementary motor area (SMA), PCL), which are related to motor function and memory. 
In the ROI-level subnetwork in (d), the positive part was the anterior part of brain including the prefrontal, temporal, and parietal regions with the ACC and CAU, and the negative part was the cerebellar and occipital regions with the PoCG. 
The positive and negative parts of the subnetworks were negatively correlated, and decreased and increased with the disease progression, respectively, because their features decreased over the groups in the first row of Fig. \ref{fig:ADNI} (a,b). 
 
\begin{figure}[t]
	\centering
	\includegraphics[width=1\linewidth]{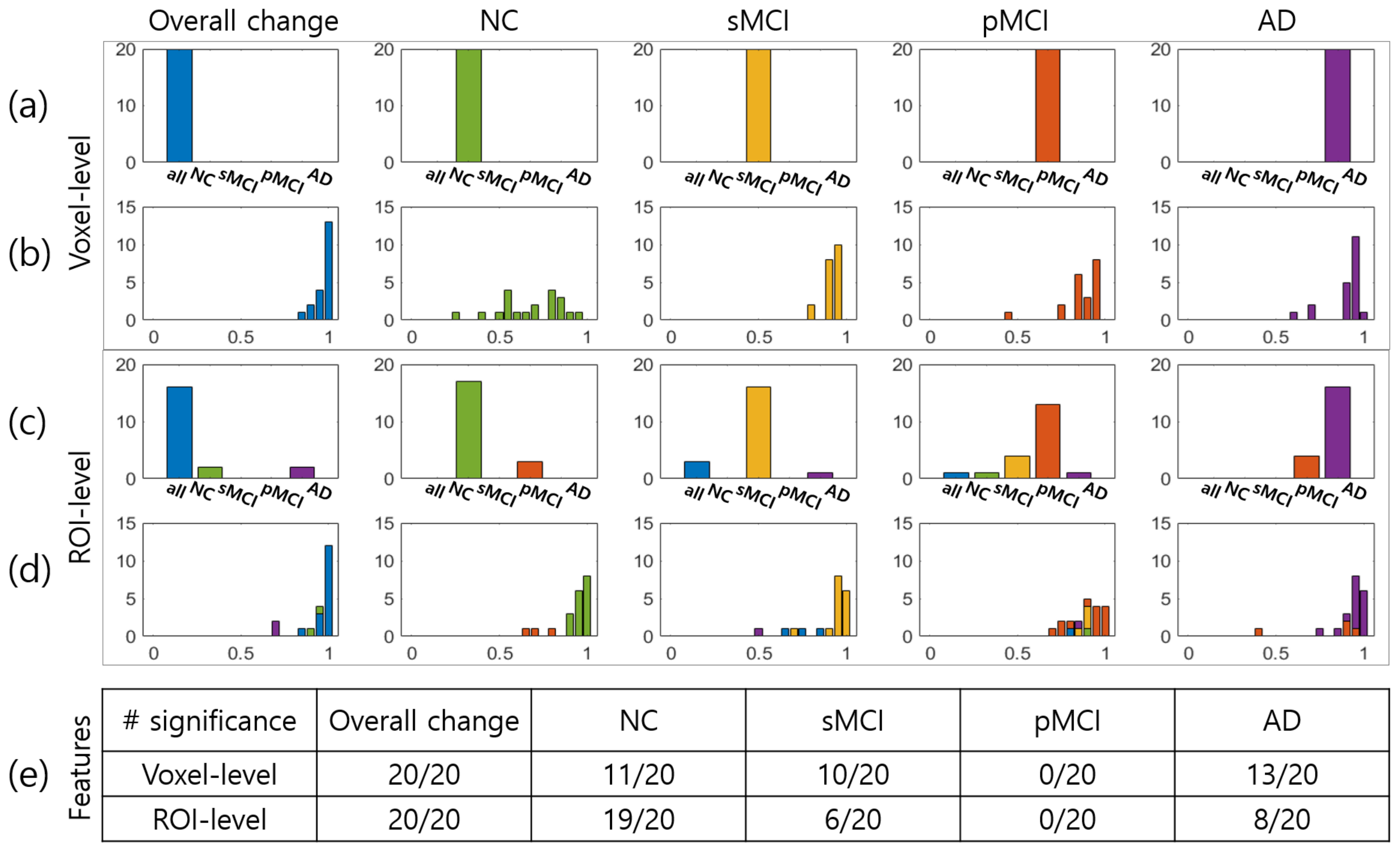}
	\caption{(a,c) Best match between the dominant spatial maps of overall change, NC, sMCI, pMCI, and AD in Fig. \ref{fig:ADNI} (a,b) and those obtained by 10 repeats of a two-fold cross validation (20 different sets). (b,d) Histograms of correlation values at the best match. The color, blue, green, yellow, red, or violet represents the spatial maps of overall change, NC, sMCI, pMCI, and AD in the 20 sets, respectively. (e) Number of times when features were significantly distinguished between groups among the 20 sets (corrected $p<.005$). }
	\label{fig:ADNI_reliability}
	\vspace{-0.3cm}
\end{figure}

\subsubsection{Reliability}
 
To check the reliability of the dominant spatial maps, corresponding features and dendrograms, we repeated a two-fold cross validation 10 times, and obtained 20 different sets for all groups, NC, sMCI, pMCI, and AD.
To assess the reliability of the spatial maps, we estimated the Pearson correlation between the dominant spatial maps of all groups, NC, sMCI, pMCI, and AD in Fig.  \ref{fig:ADNI} (a,b) and those obtained by the set, and found a best match between them using the Hungarian algorithm \citep{papadimitriou.1982.book}. 
In Fig. \ref{fig:ADNI_reliability} (a), in voxel-level analysis, the dominant spatial maps of overall change, NC, sMCI, pMCI, and AD in Fig. \ref{fig:ADNI} (a) from left to right columns are always maximally matched to those obtained by the cross validation.  
However, in Fig. \ref{fig:ADNI_reliability} (c), in ROI-level analysis, they were maximally matched to 80 \% overall, 85 \% NC, 80 \% sMCI, 65 \% pMCI, and 80 \% AD spatial maps among 20 different sets. 
The histograms of the correlation values at the best match are plotted in (b) and (d). 
In (b)-NC, the correlation values at the best match for the 20 sets were not high, with the minimum correlation value 0.27. 
However, even the minimum value was significant, because the dimension of the spatial maps is  182,363. 
In (d), the correlation values of the 20 sets tended to be high when they were best matched with the same group (blue for overall change, green for NC, yellow for sMCI, red for pMCI, and violet for AD). 

We counted the number of times that the features of overall change were significantly distinguished between the four groups, using Kruskal-Wallis tests, and those of NC, sMCI, pMCI, and AD were significantly distinguished between the group and the other groups by Wilconxon rank sum tests  (Bonferroni corrected $p<.005$). 
The results are shown in Fig. \ref{fig:ADNI_reliability} (e). 
Only the features of overall change showed the 100 \% reproducibility of discrimination for the 20 sets at both voxel- and ROI-level. 
To check the change in the structure of the original and smoothed dendrograms over the groups, we counted the number of all PCCs, all leaf bars, all layers, and voxels covered by all activated brain regions in the voxel-level original and smoothed dendrograms of the 20 sets, and the results of the group difference are summarized in Table \ref{table:reliability}. 

\begin{table}[h]
\caption{Detailed names of activated brain regions in Fig. \ref{fig:ICMs}.}
\label{table:ICMs}
\begin{tabular}{|C{3mm} |C{2mm} |C{19mm}|C{10mm}| L{89mm} |C{16mm}|}
\hline
      &   & color        & $s_{max}$ & regions & significance \\ \hline
\multirow{7}{*}{(a)} & 1 & red & 1.88 & RL: medial part of superior frontal gyrus (SFGmed), superior frontal gyrus (SFG), Rolandic operculum (ROL) & **           \\ \cline{2-6} 
    & 2 & mint & 1.65 & L: angular gyrus (ANG), supramarginal gyrus (SMG), inferior parietal gyrus (IPG) & ** \\ \cline{2-6} 
    & 3 & yellow-green & 1.49 & RL: precuneus (PCNU), anterior and middle cingulate cortex (ACC, MCC) & ** \\ \cline{2-6} 
    & 4 & mint & 1.38 & R: ANG, middle occipital gyrus (MOG), IPG  & * \\ \cline{2-6} 
    & 5 & blue & 1.04 & R: middle and inferior temporal gyrus (MTG, ITG), and the temporal pole of middle temporal gyrus (MTGp) & * \\ \cline{2-6} 
    & 6 & red & 0.95 & L: the orbital part of inferior frontal gyrus (IFGorb) & * \\ \cline{2-6} 
    & 7 & blue & 0.93 & L: MTG, ITG, MTGp, insula (INS) & * \\ \hline
\multirow{2}{*}{(b)} & 1 & blue &  -0.95 & L: MTG and ITG & *\\ \cline{2-6} 
    & 2 & green &  -0.90 & R: PCNU & *            \\ \hline
\multirow{5}{*}{(c)} & 1 & red & 1.29 & RL: ROL, ACC, paracentral lobule (PCL) & ** \\ \cline{2-6} 
    & 2 & gray & 1.08 & L: INS, SMG, precentral gyrus (PreCG), superior temporal gyrus (STG), the pars triangular of the inferior frontal gyrus (IFGtri), putamen (PUT),  & ** \\ \cline{2-6} 
    & 3 & gray & 1.07 & R: INS, SMG, PreCG, PUT, STG & ** \\ \cline{2-6} 
    & 4 & red  & 0.74       & L: middle frontal gyrus (MFG), SFG, IFGtri & **           \\ \cline{2-6} 
    & 5 & red & 0.67       & R: MFG and SFG & *            \\ \hline
\end{tabular}
\\ \\ $s_{max}:$ the maximum peak value in an activated brain region, R: right hemisphere, L: left hemisphere, **: $Z > 6,$ *: $Z>4$
\vspace{-0.2cm}
\end{table}

\begin{table}[h]
\caption{Change of the structure of voxel-level original and smoothed dendrograms during Alzheimer's disease progression (Wilcoxon rank sum test, Boferroni corrected $p<.005$).}
\label{table:reliability} 
\begin{center}
\begin{tabular}{|c|c|c|}
\hline
Original  & Positive                                & Negative                                \\ \hline
\# PCCs   & NC$\approx$sMCI$\approx$pMCI$>$AD       & NC$\approx$sMCI$\approx$pMCI$>$AD       \\
\# leaves & NC$\approx$sMCI$\approx$pMCI$>$AD       & NC$\approx$sMCI$\approx$pMCI$>$AD       \\
\# layers & NC$\approx$sMCI$\approx$pMCI$\approx$AD & NC$\approx$sMCI$\approx$pMCI$\approx$AD \\
\# voxels & NC$\approx$sMCI$\approx$pMCI$>$AD       & NC$\approx$sMCI$>$pMCI$>$AD             \\ \hline
Smoothed  & Positive                                & Negative                                \\ \hline
\# PCCs   & sMCI$>$NC$>$pMCI$>$AD                   & sMCI$\approx$NC$>$pMCI$>$AD             \\
\# leaves & sMCI$>$NC$>$pMCI$>$AD                   & sMCI$\approx$NC$>$pMCI$>$AD             \\
\# layers & sMCI$>$NC$>$pMCI$>$AD                   & sMCI$>$NC$>$pMCI$>$AD                   \\
\# voxels & NC$\approx$sMCI$\approx$pMCI$<$AD       & NC$\approx$sMCI$\approx$pMCI$\approx$AD \\ \hline
\end{tabular}
\end{center}
\vspace{-0.3cm}
\end{table}

\section{Discussion and conclusions} 

In this study, we developed a threshold-free activated region extraction method from a brain spatial map by dendrogramizing and smoothing the map. 
To verify the proposed method, we also demonstrated the relationship between Rips filtration and Morse filtration.  
The proposed method can simplify a Morse function including a brain spatial map in a three-dimensional space to a dendrogram in a two-dimensional plane. 
The dendrogramized spatial map can visualize activated brain regions as well as the range of their importance in a spatial map. 
Therefore, a user can check the shape of a brain spatial map estimated by his or her own method, and choose a proper threshold by monitoring the dendrogramized map for interpretation. 
In Sec. \ref{sec:ICAresult}, if the significance level was appropriately adjusted within the allowable range by dendrograms, the brain symmetry of DMN and SN could be preserved. 
We could also choose the threshold by using the large gaps between activated brain regions, not only by the significance level.   
The proposed method can help to identify the shape of a spatial map in a brain space, rather than in a histogram or vectorized form of the map. 
In Table \ref{table:reliability}, the smoothed dendrogram of the positive part in AD had a small number of layers and leaf bars, while there were a large number of voxels in activated brain regions. 
This finding might indicate that the function of brain regions in AD was fused and crushed rather than performing their own functions.   
Another advantage of the proposed method is the ability to transform a spatial map (function) to a network, as shown in Sec. \ref{sec:ADNIresult}. 
Since PCA and ICA are multivariate approaches, the obtained spatial map contains the information about the relationship between variables (voxels). 
The proposed method could express this hidden relationship in a spatial map as a network, as shown in Fig. \ref{fig:ADNI} (c), in which DMN was found in the positive part, and a network related with motor function and memory was found in the negative part. 

The disadvantage of the proposed method is that it can be affected by the spatial smoothing of the brain imaging data. 
When the proposed method was applied to non-smoothed data, the activated brain regions were like jagged clusters. 
Therefore, smoothing is necessary before applying the proposed dendrogramization. 
As can be seen in toy example in the Results section, the duration of the activated regions in the dendrogram was sensitive to the noise level, while the activated regions themselves were comparatively well estimated. 
Therefore, in the future, we need to develop an additional procedure to correct the duration according to the noise level.  

The proposed method can be applied to a statistical map obtained in a hypothesis-driven way. 
However, there is a risk of increasing the false positive rate. 
We therefore limited the application of the proposed method to a brain spatial map obtained by a data-driven LMF model in this study. 
However, a method to appropriately adjust the cluster size and significance level could provide a new approach to cluster-based statistics in the future.

\section*{Acknowledgement} 

Data collection and sharing for this project was funded by the Alzheimer's Disease Neuroimaging Initiative (ADNI) (National Institutes of Health Grant U01 AG024904) and DOD ADNI (Department of Defense award number W81XWH-12-2-0012). ADNI is funded by the National Institute on Aging, the National Institute of Biomedical Imaging and Bioengineering, and through generous contributions from the following: AbbVie, Alzheimer’s Association; Alzheimer’s Drug Discovery Foundation; Araclon Biotech; BioClinica, Inc.; Biogen; Bristol-Myers Squibb Company; CereSpir, Inc.; Cogstate; Eisai Inc.; Elan Pharmaceuticals, Inc.; Eli Lilly and Company; EuroImmun; F. Hoffmann-La Roche Ltd and its affiliated company Genentech, Inc.; Fujirebio; GE Healthcare; IXICO Ltd.; Janssen Alzheimer Immunotherapy Research \& Development, LLC.; Johnson \& Johnson Pharmaceutical Research \& Development LLC.; Lumosity; Lundbeck; Merck \& Co., Inc.; Meso Scale Diagnostics, LLC.; NeuroRx Research; Neurotrack Technologies; Novartis Pharmaceuticals Corporation; Pfizer Inc.; Piramal Imaging; Servier; Takeda Pharmaceutical Company; and Transition Therapeutics. The Canadian Institutes of Health Research is providing funds to support ADNI clinical sites in Canada. Private sector contributions are facilitated by the Foundation for the National Institutes of Health (www.fnih.org). The grantee organization is the Northern California Institute for Research and Education, and the study is coordinated by the Alzheimer’s Therapeutic Research Institute at the University of Southern California. ADNI data are disseminated by the Laboratory for Neuro Imaging at the University of Southern California.
This research was supported by Basic Science Research Program through the National Research Foundation of Korea (NRF) funded by the Ministry of Education (2020R1A2C1013853, 2016R1D1A1B03935, 2020R1A2C2011532, and 2017M3C7A1048079).

\section*{Appendix: Proof of theorem 1} 
\label{app:theorem1}

	\begin{enumerate} 
		\item Nonnegativity and small self-distances:  $0 \le d_{ii} \le d_{ij} = \max(d_{i},d_{j})$
		\item Indistancy implies equality: 
		if $d_{i} = d_{j} = d_{ij} = \max(d_{i},d_{j}),$ then, $v_{i}$ and $v_{j}$ are connected as soon as they appear during Morse filtration. They are the same. 
		\item Symmetry: $d_{ij}=\max(d_{i},d_{j})=d_{ji}.$
		\item Triangularity: given $v_{i},$ $v_{j},$ and $v_{k},$ 
		\begin{enumerate} 
			\item if they all are not adjacent, $d_{ij}+d_{jk}-d_{j}-d_{ik} = d_{\max} + d_{\max} - d_{j} - d_{\max} > 0.$
			\item if $v_{i}$ and $v_{j}$ are adjacent, but $v_{k}$ is not connected, $d_{ij}+d_{jk}-d_{j}-d_{ik} = \max(d_{i},d_{j})+d_{\max} - d_{j} - d_{\max} \ge 0.$
			\item if $v_{j}$ and $v_{k}$ are adjacent, but $v_{i}$ is not connected, $d_{ij}+d_{jk}-d_{j}-d_{ik} = d_{\max} +\max(d_{j},d_{k}) - d_{j} - d_{\max} \ge 0.$
			\item if $v_{i}$ and $v_{k}$ are adjacent, but $v_{j}$ is not connected, $d_{ij}+d_{jk}-d_{j}-d_{ik} = d_{\max}+d_{\max} - d_{j} - \max(d_{i},d_{k}) \ge 0.$
			\item if they all are adjacent, $d_{ij}+d_{jk}-d_{j}-d_{ik} = \max(d_{i},d_{j})+\max(d_{j},d_{k}) - d_{j} - \max(d_{i},d_{k})= \calA$. 
			\begin{enumerate}
			\item $d_{i} \ge d_{j} \ge d_{k}: \calA =d_{i} + d_{j} - d_{j} -d_{i} =0 $
			\item $d_{i} \ge d_{k} \ge d_{j}: \calA =d_{i} + d_{k} - d_{j} -d_{i} \ge 0 $
			\item $d_{j} \ge d_{i} \ge d_{k}: \calA =d_{j} + d_{j} - d_{j} - d_{i} \ge 0$ 
			\item $d_{j} \ge d_{k} \ge d_{i}: \calA =d_{j} + d_{j} - d_{j} - d_{k} \ge 0$ 
			\item $d_{k} \ge d_{i} \ge d_{j}: \calA =d_{i} + d_{k} - d_{j} - d_{k} \ge 0$ 
			\item $d_{k} \ge d_{j} \ge d_{i}: \calA = d_{j} + d_{k} - d_{j} - d_{k} = 0$
			\end{enumerate} 
		\end{enumerate} 
		Therefore, $d_{ik} \le d_{ij} + d_{jk} - d_{jj}.$
	\end{enumerate}
	The given $p$ nodes with distance $\bD$ satisfies the properties of a partial metric space.

\bibliographystyle{unsrtnat}
\bibliography{leehk}  






\end{document}